\documentclass{article}

\usepackage{epsfig}
\usepackage{subfigure}
\usepackage{calc}
\usepackage{graphicx}

\usepackage{amsfonts}
\usepackage{amsmath}
\usepackage{amssymb}
\usepackage{latexsym}

\begin{document}

\title{Likelihood that a pseudorandom sequence generator has optimal properties}
\date{}
\author{A. F\'{u}ster-Sabater$^{(1)}$ and L.J. Garc\'{i}a-Villalba$^{(1)}$\\
{\small (1) Department of Information Processing and Coding} \\
{\small Institute of Applied Physics, C.S.I.C., Serrano 144, 28006 Madrid, Spain} \\
{\small amparo@iec.csic.es}}

\maketitle

\begin{abstract}
The authors prove that the probability of choosing a nonlinear
filter of \textit{m}-sequences with optimal properties, that is,
maximum period and maximum linear complexity, tends
assymptotically to 1 as the linear feedback shift register length
increases.
\end{abstract}

\footnotetext{Work supported by CICYT (Spain) under grant
TIC95-0800. \\
Electronics Letters. Vol. 34, No. 7, pp. 646-647. April 1998. \\
Elect. Lett. Online No: 19980499, INSPEC Accession Number:
5896277}

\vspace*{0.9cm}

\noindent Pseudorandom sequence generators have multiple
applications in radar systems, simulation, error-correcting codes,
spread-spectrum communication systems and cryptography. One of the
most interesting pseudorandom sequence generators is the nonlinear
filter of \textit{m}-sequences, as it produces sequences with
optimal properties.

A nonlinear filter $F$ is a \textit{k}th order nonlinear function
applied to the $L$ stages of an LFSR with a primitive feedback
polynomial. Let $\{a_n\}$ be the LFSR output sequence; then the
generic element $a_n$ is $a_n=\alpha^n + \alpha^{2n}+ ... +
\alpha^{2^{(L-1)}n}$, $\alpha \in GF(2^L)$ being a root of the
LFSR characteristic polynomial. Thus, the filtered sequence
$\{z_n\}$ can be represented as

\[\{z_n\}=\{F(a_n,\cdots, a_{n+L-1})\}
\]

\[= \sum\limits_{i=1}^{N} \{C_i \alpha^{E_in}+ \cdots + ( C_i\alpha^{E_in} )^{2^{(r_i-1)}}
\}= \sum\limits_{i=1}^{N} C_i \{S_n^{E_i}\}
\]

\noindent with $r_i$ being the cardinal of coset $E_i$
\cite{RUEPPEL}, $N$ the number of cosets $E_i$ with binary weight
$\leq k$ and $C_i\in GF(2^L)$ constant coefficients. Note that the
\textit{i}th term in the expression of $\{z_n\}$ corresponds to
the characteristic sequence $\{S_n^{E_i}\}$ of coset $E_i$.
Therefore $\{z_n\}$ can be written as the termwise sum of the
characteristic sequences associated with every coset $E_i$. From
the above the following can be noted:
\begin{description}
\item  (i) It can be proved \cite{LIDL} that every coefficient
    $C_i \in GF(2^{r_i})$, so that as long as $C_i$ is within
    its corresponding field, we shift along the sequence
    $\{S_n^{E_i}\}$.
\item  (ii) If $C_i = 0$, then coset $E_i$ does not contribute
    to the linear complexity of the filtered sequence
    $\{z_n\}$.
\item  (iii) The period of $\{z_n\}$ is the minimum common
    multiple of the periods of its corresponding
    characteristic sequences $\{S_n^{E_i}\}$ whose values are
    the divisors of $2^{L} - 1$. \end{description}

Taking the above considerations into account, we can compute the
probability of choosing a nonlinear filter $F$, whose output
sequence $\{z_n\}$ has optimal properties. In fact, let $nfk$ be
the number of \textit{k}th order nonlinear filter functions and
$nfm$ the number of the previous functions whose output sequences
$\{z_n\}$ have maximun linear complexity $(C_i\neq 0, \forall i)$,
then
\[ Pr=\frac{nfm}{nfk}=\frac{(2^{r_1-1}-1)\;(2^{r_2-1}-1)\cdots (2^{r_N-1}-1)}{(2^{L \choose k}-1)\; 2^{L \choose {k-1}}\cdots \;2^{L \choose 1}}
\]

\[=\frac{\prod\limits_{i=1}^N \; (2^{r_i-1}-1)} {(2^{L \choose k}-1)\; 2^{L \choose {k-1}}\cdots \;2^{L \choose 1}}\]
If $L$ is prime (which is the most common case), then all the
cardinals $r_i$ equal $L$. Consequently, $nfm$ and $Pr$ can be
rewritten as

\[ nfm = (2^{L} - 1)^N = (2^{L} - 1)^{ \frac{1}{L}  \;  \sum\limits_{i=1}^{k}   {L \choose k}} = (2^{L} -
1)^{\frac{N_k}{L}}
\]

\[ Pr=\frac{(2^{L} -
1)^{\frac{N_k}{L}}}{(2^{L \choose k}-1)\; 2^{L \choose {k-1}}\cdots \;2^{L \choose 1}}
\]

\[ > \frac{(2^{L} -
1)^{\frac{N_k}{L}} } {2^{N_k}} = \Big(\frac{2^{L} -1}{2^{L}}\Big)^{\frac{N_k}{L}} = \Big(1 - \frac{1} {2^L}\Big)^{2^{L}\; \frac{N_k} {2^{L}\;L}}
\]

It is a well known fact that if $b_n \rightarrow \infty $, then
$(1-b_n^{-1})^{b_n} \rightarrow e^{-1}$. As $N_k\leq 2^{L}-1$, if
$k \simeq L/2$ then $N_k \simeq 2^{L-1}$. Thus,

\[ Pr> e^{- \frac{N_k} {{2^{L}\;L}} } \simeq e^{- \frac{1} {2L} }
\]

For $L = 257$ (a typical value for the LFSR in communication
systems), $Pr > 0.998$

In addition, this kind of nonlinear filter also has maximum
period. Indeed, as those filters contain the characteristic
sequences $\{S_n^{E_i}\}$ associated with all the cosets $E_i$,
they also contain that of coset $E_1$ the period \cite{PARK} of
which is $2^{L} - 1$.

\vspace*{0.4cm}

\textit{Conclusions:} Nonlinear filters of \textit{m}-sequences
are believed to be excellent pseudorandom sequence generators.
This is not only because they are very easy to implement with
high-speed electronic devices, but also because they are highly
likely to produce sequences with optimal properties.

\vspace*{0.4cm}

\textit{Acknowledgment:} This work is supported by CICYT (Spain)
under grant TIC95-0800.

\bibliographystyle{amsalpha}

\end{document}